\documentclass[sigconf,nonacm]{acmart}
\usepackage{gensymb}
\usepackage{todonotes}
\usepackage{enumitem}
\usepackage{listings}
\usepackage{algorithm,float}
\usepackage{subcaption}
\lstdefinelanguage{ttl} {
language={SQL},
    alsoletter={-},
    morekeywords={bif,skos,owl,dbr,rdf,sem,so,uner,oekg-s,wkg,wkgs,geo,rdfs,a,sf,uom, wd, RAND},
    deletekeywords={date}
}

\newcommand{\voc}[2]{\texttt{#1:\allowbreak #2}}
\newcommand\schema[1]{{\normalfont\fontfamily{cmvtt}\selectfont #1}}

\AtBeginDocument{%
  \providecommand\BibTeX{{%
    \normalfont B\kern-0.5em{\scshape i\kern-0.25em b}\kern-0.8em\TeX}}}

\copyrightyear{2021}
\acmYear{2021}
\setcopyright{acmlicensed}\acmConference[CIKM '21]{Proceedings of the 30th ACM International Conference on Information and Knowledge Management}{November 1--5, 2021}{Virtual Event, QLD, Australia}
\acmBooktitle{Proceedings of the 30th ACM International Conference on Information and Knowledge Management (CIKM '21), November 1--5, 2021, Virtual Event, QLD, Australia}
\acmPrice{15.00}
\acmDOI{10.1145/3459637.3482023}
\acmISBN{978-1-4503-8446-9/21/11}

\begin{document}
\title[WorldKG]{WorldKG: A World-Scale Geographic Knowledge Graph}



\author{Alishiba Dsouza}
\email{dsouza@cs.uni-bonn.de}
\affiliation{%
  \institution{Data Science \& Intelligent Systems\\ University of Bonn}
  \country{}
}

 \author{Nicolas Tempelmeier}
\email{tempelmeier@L3S.de}
\affiliation{%
  \institution{L3S Research Center\\Leibniz Universit\"at Hannover}
  \country{}
}

\author{Ran Yu}
\email{ran.yu@uni-bonn.de}
\affiliation{%
  \institution{Data Science \& Intelligent Systems\\University of Bonn}
  \country{}
}

\author{Simon Gottschalk}
\email{gottschalk@L3S.de}
\affiliation{%
 \institution{L3S Research Center\\Leibniz Universit\"at Hannover}
 \country{}
}

\author{Elena Demidova}
\email{elena.demidova@cs.uni-bonn.de}
\affiliation{%
  \institution{Data Science \& Intelligent Systems\\University of Bonn}
 \country{}
  }

\renewcommand{\shortauthors}{Alishiba Dsouza et al.}

\begin{abstract}
OpenStreetMap is a rich source of openly available geographic information. However, the representation of geographic entities, e.g., buildings, mountains, and cities, within OpenStreetMap is highly heterogeneous, diverse, and incomplete. As a result, this rich data source is hardly usable for real-world applications. This paper presents WorldKG - a new geographic knowledge graph aiming to provide a comprehensive semantic representation of geographic entities in OpenStreetMap. We describe the WorldKG knowledge graph, including its ontology that builds the semantic dataset backbone, the extraction procedure of the ontology and geographic entities from OpenStreetMap, and the methods to enhance entity annotation.
We perform statistical and qualitative dataset assessment, demonstrating the large scale and high precision of the semantic geographic information in WorldKG.
\end{abstract}



\settopmatter{printfolios=true}
\maketitle

\noindent\textbf{Resource type:} Dataset\\
\textbf{Website and documentation:} \url{http://www.worldkg.org} \\
\textbf{Dataset DOI:} \url{https://zenodo.org/record/4953986} \\ 

\section{Introduction}
\label{sec:introduction}

OpenStreetMap (OSM) is a rich source of openly available volunteered geographic information,
including over 6.8 billion geographic entities in 188 countries 
contributed by over 7.6 million volunteers \cite{osmstats}. 
OSM is adopted in a variety of real-world applications on the Web and beyond, including map tile generation \cite{4653466} and routing \cite{epub34045}. However, representations of geographic entities in OSM are highly diverse, including few mandatory properties and numerous heterogeneous tags, i.e., user-defined key-value pairs. 
The tag-based structure of OSM data does not follow a well-defined ontology, significantly limiting automatic interpretation and use of OSM data in real-world applications. 

Knowledge graphs (KGs) have recently emerged as a key technology to provide semantic machine-interpretable information on real-world entities at scale. However, popular general-purpose knowledge graphs such as DBpedia and Wikidata lack coverage of geographic entities \cite{TEMPELMEIER2021349}. In contrast, specialized geographic knowledge graphs such as LinkedGeoData \cite{auer2009linkedgeodata} and YAGO2geo \cite{karalis2019extending} lack coverage of geographic classes. To provide a comprehensive source of semantic geographic information at scale, semantic information in knowledge graphs and community-created geographic sources such as OSM should be tightly integrated and fused.

Integration of OSM and knowledge graphs is inherently difficult. 
Although some community-defined links between OSM entities and knowledge graphs like Wikidata exist at the instance level, these links are still sparse and cover only certain entity types. As of April 2021, only 0.52\% of OSM entities provided links to Wikidata. 
In our previous work, we proposed initial approaches for the integration of OSM and knowledge graphs at the schema \cite{iswcNCA}, and instance levels \cite{TEMPELMEIER2021349}. In this work, we build upon the Neural Class Alignment (NCA) approach \cite{iswcNCA} to provide semantic annotations to OSM entities.  
Overall, further research efforts are required to facilitate tighter integration and fusion of OSM and knowledge graphs.

This paper presents WorldKG -- a novel comprehensive geographic knowledge graph built from the OSM dataset. 
We create a novel WorldKG ontology by converting the flat OSM schema into a hierarchical ontology structure. 
The current release of WorldKG V1.0 in June 2021 contains over 100 million geographic entities from 188 countries and over 800 million triples.
Overall, the number of geographic entities in WorldKG is two orders of 
magnitude higher than in Wikidata and DBpedia knowledge graphs.
To facilitate the adoption of WorldKG in semantic applications, 
we align the WorldKG ontology with the Wikidata and DBpedia ontologies using the NCA approach proposed in our previous work \cite{iswcNCA}.
Our evaluation results demonstrate that the alignment enables us to determine correct Wikidata and DBpedia ontology classes of WorldKG entities with over 99\% accuracy, on average.

The scale and accuracy of WorldKG can facilitate the broader adoption of semantic geographic knowledge in a variety of real-world applications. Examples include event-centric \cite{DBLP:conf/cikm/CostaGD20} and geospatial \cite{punjani2018template} question answering, geographic information retrieval \cite{INR-034}, and other cross-domain semantic data-driven applications.

Overall, our main contributions in this paper are as follows:

\begin{itemize}
    \item We present WorldKG -- a new knowledge graph containing large-scale semantic geographic data extracted from OSM. 
    \item We present the WorldKG ontology, which semantically describes geographic entities and links them to the specific classes in the Wikidata and DBpedia ontologies.
    \item We provide access to WorldKG through a SPARQL endpoint and provide downloadable data files in the standard RDF turtle format~\cite{turtle2014}.
    \item To ensure reproducibility, we make the source code of the whole pipeline for WorldKG creation publicly available on GitHub under an open MIT license.
\end{itemize}

The rest of the paper is organized as follows: In Section \ref{sec:relevance}, 
we discuss the relevance and the expected impact of the proposed WorldKG knowledge graph. 
Then, we provide formal definitions of an OSM corpus and a knowledge graph in Section \ref{sec:OSMandKG}. We introduce the proposed WorldKG ontology in Section \ref{sec:ontology} and explain the WorldKG creation process in Section \ref{sec:cgeneration}. We present the statistics and evaluation results of WorldKG in Section \ref{sec:evaluation}. In Section \ref{sec:availaility}, we describe the availability, utility, and sustainability aspects of our dataset. 
Section \ref{sec:example} provides a real-world application example using WorldKG. We discuss related work in Section \ref{sec:related}. Finally, in Section \ref{sec:conclusion}, we provide concluding remarks.

\section{Relevance and Expected Impact}
\label{sec:relevance}

This section discusses the expected impact of the proposed WorldKG knowledge graph and its significance to the community, applications, and technology adoption. 

\textit{Relevance to the information and knowledge management community.} 
Large-scale volunteered geographic information has facilitated many widely used applications such as routing services and data visualizations\footnote{OSM-based services: \url{https://wiki.openstreetmap.org/wiki/List_of_OSM-based_services}}. Nevertheless, due to the data heterogeneity, the potential of such collectively created knowledge is not yet fully exploited. 
By integrating heterogeneous OSM data using semantic technologies, we construct and maintain a large-scale knowledge graph that consistently represents geographic data originating from different sources and links this data to the relevant entity types in cross-domain knowledge graphs. WorldKG constitutes a geographic data source of semantic representations with high connectivity, interoperability, and accessibility. 
In the Semantic Web community context, WorldKG provides richer information of geographic entities than the existing cross-domain knowledge graphs. Thus, WorldKG can support the development of various applications, including geographic question answering and information retrieval, point of interest recommendation, and other cross-domain semantic data-driven applications. 

\textit{Relevance for OpenStreetMap applications.} Currently, routing and navigation services such as Useful Maps 2\footnote{Useful Maps 2: \url{https://map.atownsend.org.uk/maps/map/map.html}} and Baidu Maps\footnote{Baidu Maps: \url{http://j.map.baidu.com/1CWxF}}, and visualization tools based on geographic information (e.g., weather map\footnote{Weather map: \url{https://maps.darksky.net/}}) are utilizing OSM. Meanwhile, geographic entities in cross-domain knowledge graphs have been used for entity relation referencing, question answering, and other tasks. 
However, on the one hand, OSM lacks contextual information on its nodes; on the other hand, cross-domain knowledge graphs are not well-populated with up-to-date geographic information. For these reasons, the gaps between OSM and knowledge graphs persist, and the potential of applications utilizing either type of information is substantially limited. 
By linking OSM nodes to the classes and entities in cross-domain knowledge graphs, WorldKG provides rich contextual information of the geographic entities, which can be used to enhance the existing services. For instance, enriching maps can provide more detailed location information and interconnect different information types (e.g., locations, weather, and events). 

\textit{Impact on the adoption of Semantic Web technologies.} By following best practices in data publishing and maintenance, we ensure the availability and the extensibility of WorldKG. By adopting Semantic Web technologies and standards, the accessibility and reusability of OpenStreetMap data are largely improved, and the effort associated with reusing this data is reduced significantly. With a commitment to maintaining regular updates, we ensure the sustainability of WorldKG. 
We believe that WorldKG can benefit researchers in various research fields. Examples include geographic and semantic data management, geographic information retrieval, and recommendation. Furthermore, WorldKG can accelerate the development and enhancement of various services, including interactive maps, smart assistants, and geographic recommender systems.

\section{OSM and Knowledge Graphs}
\label{sec:OSMandKG}
WorldKG targets the integration of OpenStreetMap and knowledge
graphs. In this section, we briefly describe both data structures and their interlinking.
In the context of this work, we refer to the entities with geographic extent,
i.e., the entities located on the globe, as geographic entities.

\subsection{OpenStreetMap}
OpenStreetMap is one of the essential sources of openly available volunteered geographic information globally, including contributions from over 7.6 million volunteers (as of June 2021). 
OSM captures a vast and continuously growing number of geographic entities, currently counting more than 6.8 billion in 188 countries \cite{osmstats}. 
The essential components of the OSM data model are nodes, ways, and relations.
Nodes represent entities with a geographic point location (e.g., mountain peaks and trees).
Ways represent geographic entities of a linear form (e.g., rivers and roads).
Relations are groups of elements consisting of nodes, ways, and other relations (e.g., boundaries and bus routes).
For the current scope of the WorldKG knowledge graph (WorldKG V1.0), we only consider OSM nodes. 

OSM does not follow a strict schema but provides a set of guidelines\footnote{OSM ``How to map a'': \url{https://wiki.openstreetmap.org/wiki/How_to_map_a}} for volunteers to create and annotate geographic entities.
As a result, OSM has a rich and diverse schema with over 80 thousand distinct keys and numerous values.

We formally define the concept of an OSM corpus as follows:

\begin{definition}
\label{def:osm}
An \emph{OSM corpus} $\mathcal{C}  = (N, T)$ 
consists of a set of nodes $N$ representing geographic entities, and a set of tags $T$. 
Each tag $t \in T$ is represented as a key-value pair, with the key $k \in K$ and a value $v \in V$: $t=\langle k,v \rangle$. 
A node $n \in N$,  $n=\langle i, l, T_{n} \rangle$ is represented as a tuple containing an identifier $i$, a geographic location $l$, and a set of tags $T_{n}\subset T$. 
\end{definition}

OSM nodes have a unique identifier and contain various key-value pairs called tags. The following example of the ``Zugspitze'', the highest mountain of Germany, illustrates the tag structure.

\begin{center}
$
\begin{array}{ll}
\hline
\textbf{Key} &  \textbf{Value}\\
\hline
    id & 27384190 \\
    \texttt{name}& Zugspitze \\
    \texttt{natural} & peak \\
    \texttt{summit:cross} & yes\\
    \texttt{ele} & 2962\\\\
\end{array}
$
\end{center}

Here, the tags with keys such as \textit{summit:cross}, \textit{name} and \textit{ele} (elevation above sea level) serve as properties of the entity, whereas the tag \texttt{natural=peak} represent the entity type (in this case equivalent to the DBpedia class \voc{dbo}{Mountain}). 

\subsection{Knowledge Graphs}
Knowledge graphs are a rich source of semantic information, containing entities, classes, properties, literals, and relations.

\begin{definition}
A \emph{knowledge graph} $\mathcal{KG}= (E, C, P, L, F)$
consists of a set of entities $E$, a set of classes $C \subset E$, a set of properties $P$, a set of literals $L$,
and a set of relations $F \subseteq E \times P \times (E \cup L)$. 
\end{definition}
Entities in $E$ represent real-world entities and semantic classes.
In the context of this work, we are particularly interested in geographic entities in a knowledge graph.
Properties in $P$ represent relations connecting two entities, or an entity and a literal value.
An entity in a KG can belong to one or more classes, and is typically linked to a class using \texttt{rdf:type} or an equivalent property.

\begin{definition}
The class of the entity $e\in E$ in the knowledge graph $\mathcal{KG}= (E, C, P, L, F)$ is denoted as: $\textit{class(e)} = \allowbreak \{ c \in C ~| \allowbreak ~(e, \texttt{rdf:type}, \allowbreak c) \in F \}$.
\end{definition}

The data in a knowledge graph is typically represented in the RDF\footnote{RDF: \url{https://www.w3.org/RDF/}} format having a \textit{subject - predicate - object} structure.
Consider the corresponding excerpt from the representation of the entity 
``Zugspitze'' in Wikidata:\\

\begin{center}

$
\begin{array}{lll} 
\hline
\textbf{Subject} &  \textbf{Predicate} & \textbf{Object}\\
\hline
    Q3375 & label & Zugspitze \\
    Q3375 & instance ~of & mountain \\
    Q3375 & coordinate &  47\degree25'N, 10\degree59'E \\
    Q3375 & parent peak & Q15127 \\\\

\end{array}
$
\end{center}

In this example, the statement ``Q3375 instance of mountain'' indicates that the entity belongs to the Wikidata class ``mountain''.

\subsection{Linking OpenStreetMap and KGs}

Although OSM contains a vast amount of geospatial data, OSM keys and tags are heterogeneous, do not possess any machine-readable semantics, and are not directly accessible for semantic applications. Knowledge graphs such as Wikidata, DBpedia, and YAGO provide rich ontologies but lack geographic coverage. For instance, in June 2021, 931,574 nodes with the tag \texttt{amenity=restaurant} were present in OSM, whereas Wikidata included only 4,391 entities for the equivalent class ``restaurant''. 

Equivalence links between OSM tags and knowledge graph classes are rarely present. Out of around 80,000 OSM keys, only 0.7\% are mapped to Wikidata classes. At the ontology level, the alignment is limited by the structural mismatch between the flat OSM schema and hierarchical KG ontologies. Due to the reasons above, the fusion of OSM and KG entities to create a comprehensive semantic geospatial resource is a challenging task.

\section{WorldKG Ontology}
\label{sec:ontology}
The purpose of WorldKG is to provide a comprehensive geospatial knowledge graph by integrating various data sources.
We consider the following goals while building the WorldKG ontology:
\begin{itemize}
    \item To capture geospatial entities in WorldKG.
    \item To include relations between classes of existing knowledge graphs and  WorldKG classes.
    \item To lift the OSM schema into a hierarchical ontology.
    \item To provide provenance information for all WorldKG entities.
    \item To allow for easy extensions of the WorldKG ontology.
\end{itemize}

\begin{table}
\caption{List of prefixes and namespaces used by WorldKG.}

\begin{tabular}{ll}
\toprule
\textbf{Prefix} & \textbf{Namespace} \\
\midrule
dcterms   & \url{http://purl.org/dc/terms/} \\ 
geo       & \url{http://www.opengis.net/ont/geosparql\#}       \\ 
osmn   &   \url{https://www.openstreetmap.org/node/} \\ 
owl       & \url{http://www.w3.org/2002/07/owl#} \\ 
rdf       & \url{http://www.w3.org/1999/02/22-rdf-syntax-ns#}\\ 
rdfs      & \url{http://www.w3.org/2000/01/rdf-schema\#}      \\ 
sf        & \url{http://www.opengis.net/ont/sf\#}             \\ 
uom & \url{http://www.opengis.net/def/uom/OGC/1.0/} \\
wd        & \url{http://www.wikidata.org/wiki/}                \\ 
wkg       & \url{http://www.worldkg.org/resource/} \\ 
wkgs      & \url{http://www.worldkg.org/schema/}   \\ 
\bottomrule
\end{tabular}
\label{tab:prefixList}
\end{table}

\begin{figure*}
    \centering
    \includegraphics[width = 0.7\textwidth]{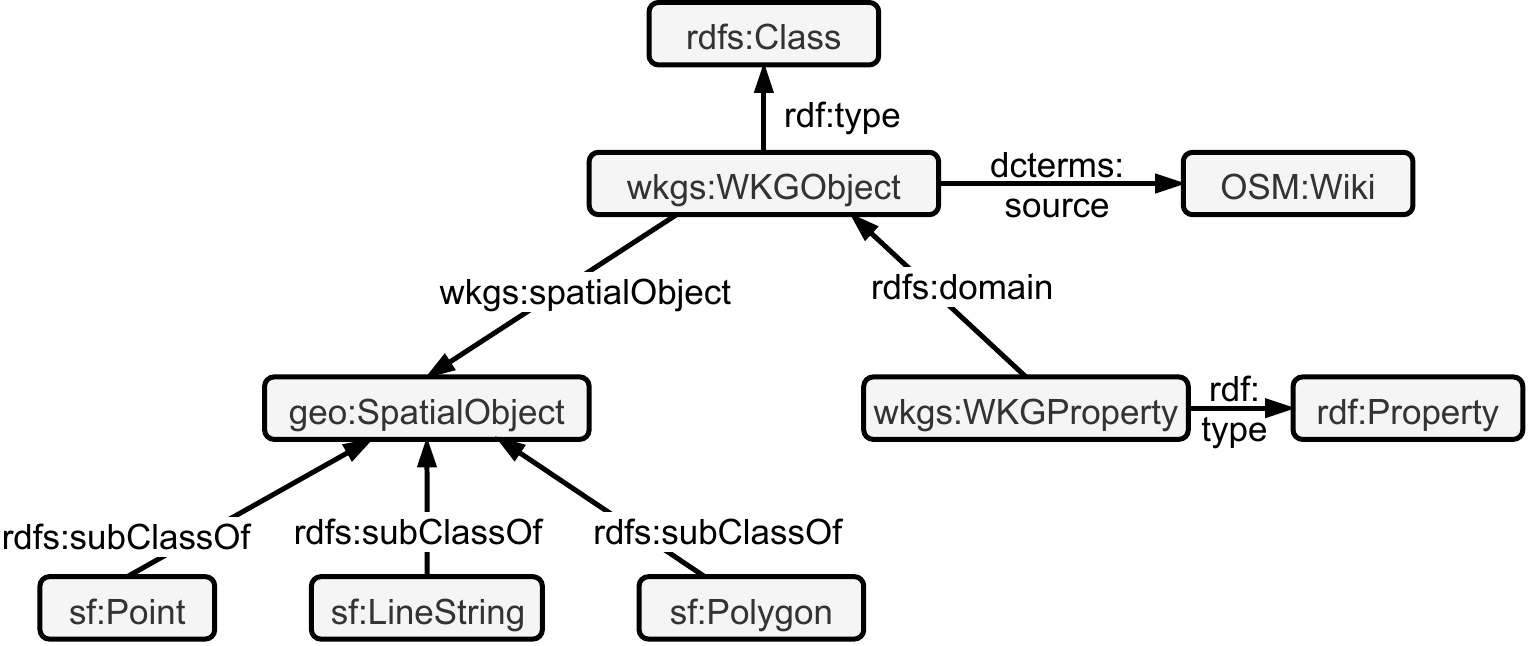}
    \caption{The WorldKG Ontology.}
    \label{fig:ontology}
\end{figure*}

We define the WorldKG ontology based on key-value pairs of the OSM schema. 
Figure \ref{fig:ontology} presents the WorldKG ontology.
Each class in the WorldKG ontology is a subclass of \voc{wkgs}{WKGObject}, where the namespace \schema{wkgs} represents WorldKG schema elements (for a list of prefixes and namespaces in WorldKG, see Table~\ref{tab:prefixList}). 
WorldKG properties are modeled as \voc{wkgs}{WKG\-Property} and provide information on OSM tags that do not indicate a type assignment.

\textit{Geospatial support.}
To enable geographic queries on the dataset, we utilize the GeoSPARQL framework proposed by the Open Geospatial Consortium\footnote{GeoSPARQL: \url{https://www.ogc.org/standards/geosparql}}.
To provide information about its geographic extent, each \voc{wkgs}{WKGObject} entity can be related to a \voc{geo}{Spatial\-Object} via the property \voc{wkgs}{spatial\-Object}, where a \voc{geo}{Spa\-ti\-al\-Ob\-ject} can be a point, line string or polygon.
\voc{geo}{Spatial\-Object} enables the computation of geospatial functions in SPARQL queries (e.g., distance, nearest neighbors).
For an example of a query using these functions, see Section~\ref{sec:example}. 

\subsection{WorldKG Classes and Properties}

The OSM community provides a list of established key-value pairs as the so-called map feature list\footnote{OSM map feature list: \url{https://wiki.openstreetmap.org/wiki/Map_features}}.
An example of a map feature is the key-value pair \texttt{natural=cave\_entrance} used to annotate cave entrances in OSM.
We use the map feature list to construct a class hierarchy.
In particular, we consider all keys in the feature map list as top-level classes (e.g., \texttt{natural}). Values assigned to the keys are represented as their subclasses. For example, \texttt{cave\_entrance} is a subclass of \texttt{natural}.

\begin{figure*}
    \centering
    \includegraphics[width = 0.7\textwidth]{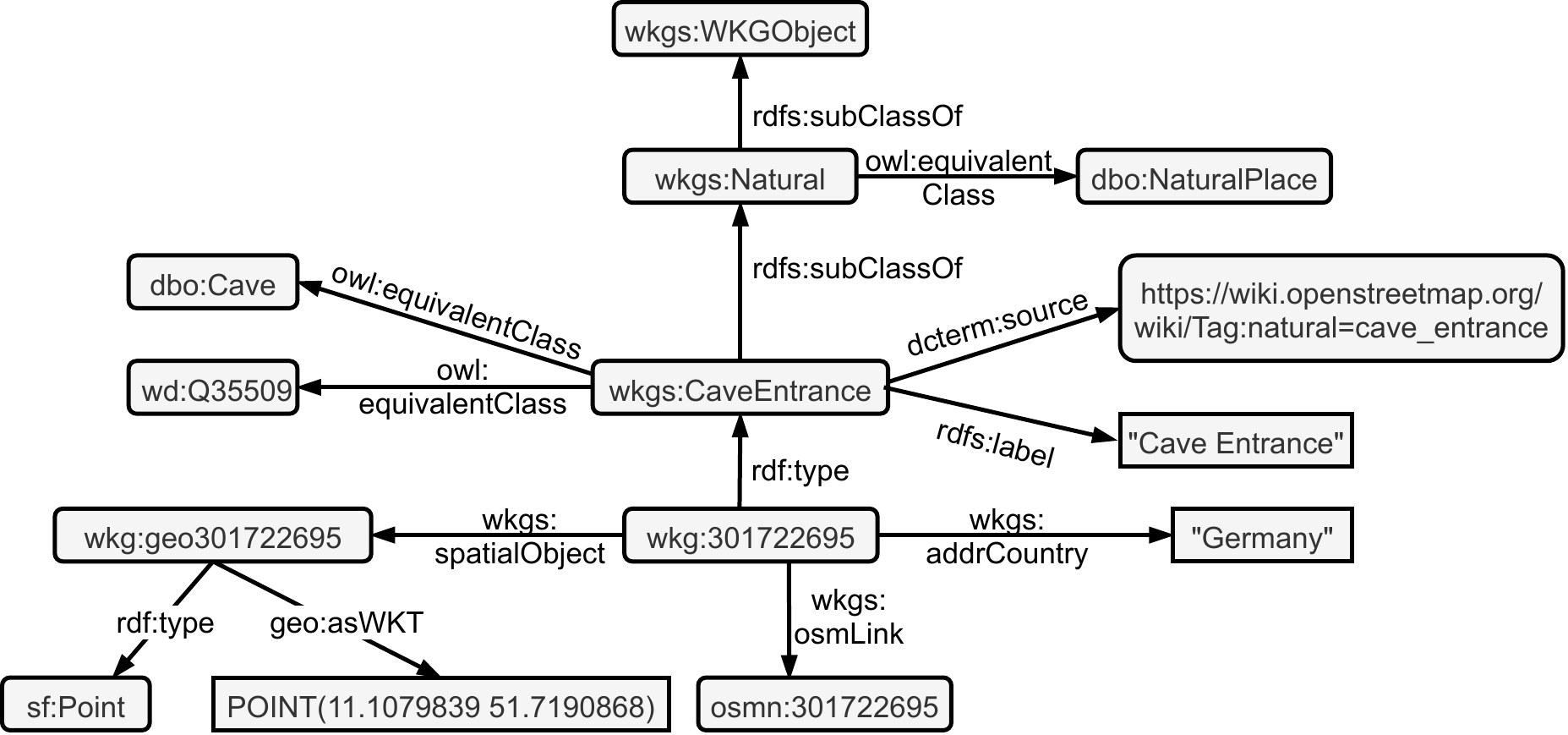}
\caption{Example instantiation of the WorldKG ontology for a specific instance of \voc{wkgs}{CaveEntrance}.}
    \label{fig:exampleNatural}
\end{figure*}

Figure \ref{fig:exampleNatural} illustrates how the key-value pair \texttt{natural=\allowbreak cave\_\allowbreak entrance} is represented in the WorldKG ontology.

\begin{itemize}
\item The OSM key \texttt{natural} is converted into the top-level class \voc{wkgs}{Natural}, which summarizes nature entities.
\item The OSM value \texttt{cave\_\allowbreak entrance} is a subclass of \voc{wkgs}{Natu\-ral}, namely \voc{wkgs}{Cave\-Entrance} representing cave entrances.
\end{itemize}

\begin{table}
\caption{Example mappings between OSM tags and Wikidata classes.}
\begin{tabular}{ll}
\toprule
\textbf{OSM tag}        & \textbf{Wikidata class (English label)} \\ \midrule
\texttt{natural=peak}       & Q8502 (mountain)          \\ 
\texttt{natural=saddle}     & Q133056 (mountain pass)       \\ 
\texttt{railway=halt}       & Q55678 (railway stop)         \\ 
\texttt{railway=station}    & Q55488 (railway station)         \\ 
\texttt{railway=tram\_stop} & Q22808404 (station located on surface)      \\ 
\texttt{building=church}     & Q16970 (church building)         \\ 
\bottomrule
\end{tabular}

\label{tab:mappingOSMKG}
\end{table}

We only consider categorical values as subclasses in WorldKG.
Other value types, e.g., boolean or numerical values, are not considered as a subclass. 
Instead, we use the top-level class provided by the corresponding key.
For example, an entity with a tag \textit{building=yes} is typed as \voc{wkgs}{Building}.

We create the properties from OSM keys that have a valid English OSM Wiki page\footnote{OSM Wiki: \url{https://wiki.openstreetmap.org/wiki/Main_Page}} and are not mapped to own classes.
In the example given in Figure \ref{fig:exampleNatural}, \voc{wkgs}{addrCountry} is inferred from a key that provides the country in which an entity is located. Each class and property is linked to an OSM Wiki page via \voc{dcterms}{source}. 

\subsection{Schema Alignment with Existing KGs}
\label{sec:schema-alignment}

To link the WorldKG ontology to other existing ontologies, we determine equivalent OSM tags and classes of the Wikidata and DBpedia knowledge graphs.
We utilize the Neural Class Alignment (NCA) approach proposed in our previous work \cite{iswcNCA} to obtain the alignments between OSM tags and the classes of established knowledge graphs. 
NCA is a 2-step unsupervised machine learning approach.
In the first step, we train a supervised neural classification model that learns to classify OSM entities into the respective knowledge graph classes based on their tags (i.e., keys and values).
After the training process is completed, we probe the resulting classification model with one tag at a time and get the class activation from the model output layer. 
Finally, we link the class and tag combinations for which the class activation exceeds an acceptance threshold $th_a$.
The detailed description of the NCA approach is provided in \cite{iswcNCA}.

We train individual models for Wikidata and DBpedia knowledge graphs. We set the acceptance threshold of NCA at $th_a=0.25$ and $th_a=0.4$ for Wikidata and DBpedia, respectively. To ensure the quality of tag-to-class matches in WorldKG, we manually verify the resulting matches and discard any wrongly mapped pairs.
Table \ref{tab:mappingOSMKG} shows example mappings between OSM tags and Wikidata classes obtained using this approach.
The alignments between the WorldKG classes and the Wikidata and DBpedia classes are represented using the \voc{owl}{equivalent\-Class} property as shown in the Figure \ref{fig:exampleNatural}.

\subsection{Geographic Entity Example}
Listing \ref{lst:exampleResTTL} illustrates an example entity description file in \textit{.ttl} format. It contains type information (\voc{wkgs}{Restaurant}) and various properties, including its label and opening hours. Via the property \voc{wkgs}{spatial\-Object}, the entity is linked to its respective \voc{geo}{Spatial\-Object}. The \voc{geo}{Spatial\-Object} represents the type of geometry of the entity (\voc{sf}{Point}) and the coordinates of the geometry. For each entity, we also provide the property \voc{wkgs}{osmLink} that links the entity to the original OSM node.

\begin{lstlisting}[float=ht,captionpos=b, caption=RDF Triples in the Turtle format for an example geographic entity of type \voc{wkgs}{Restaurant} in WorldKG., label=lst:exampleResTTL, frame=single,showstringspaces=false]
wkg:1014675277 a wkgs:Restaurant;
   rdfs:label "Krishna" ;
   wkgs:addrCountry "DE" ;
   wkgs:addrHousenumber "53;54" ;
   wkgs:cuisine "indian" ;
   wkgs:dietVegetarian "yes";
   wkgs:openingHours "Mo-Su 17:00-23:00" ;
   wkgs:organic "only" ;
   wkgs:phone "+49 421 52279939" ;
   wkgs:spatialObject wkg:geo1014675227 ;
   wkgs:website "http://www.indisches-
                     bio-restaurant.de/" ;
   wkgs:wheelchair "no" ;
   wkgs:osmLink osmn:1014675277.
   
wkg:geo1014675227 a sf:Point;
   geo:asWKT "Point(8.7938916 53.073794)"
                        ^^geo:wktLiteral .
\end{lstlisting}

\section{WorldKG Creation Process}
\label{sec:cgeneration}

\begin{figure*}[!h]
    \centering
    \includegraphics[width=0.8\textwidth]{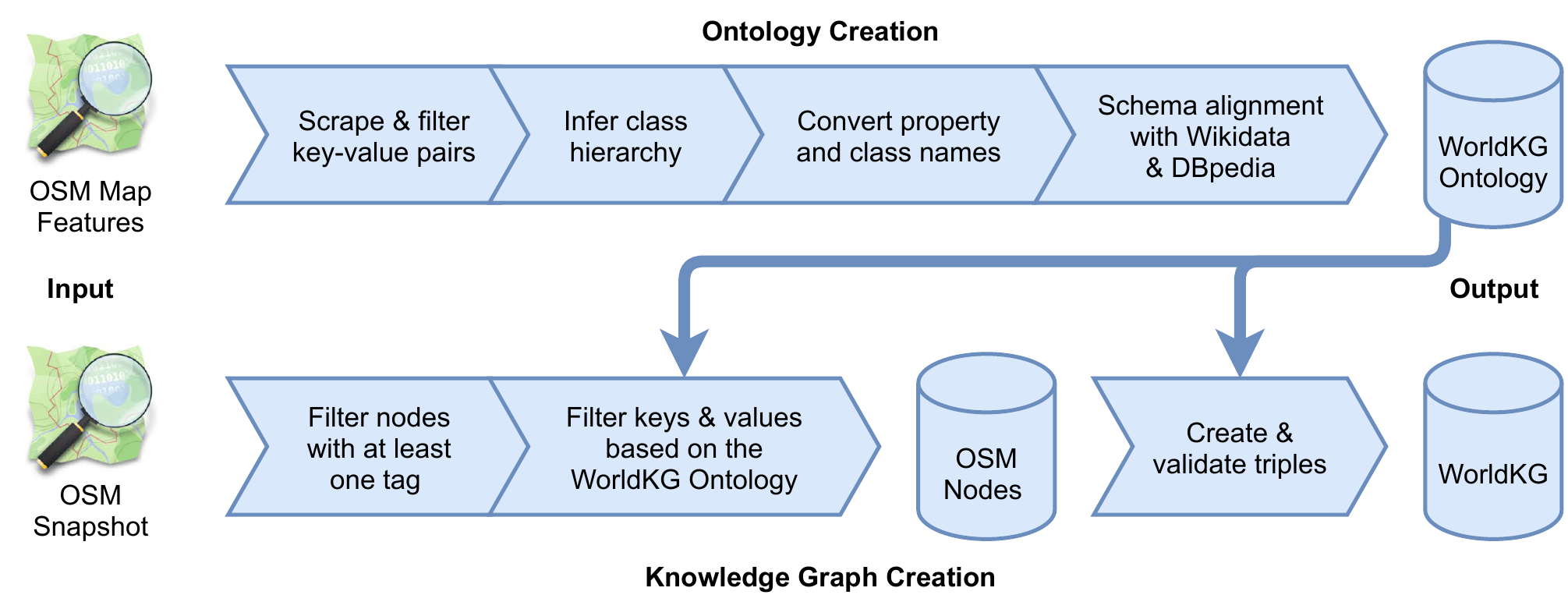}
    \caption{WorldKG ontology and knowledge graph creation process.}
    \label{fig:pipeline}
\end{figure*}

In this section, we present our approach for creating WorldKG, consisting of the WorldKG ontology and geographic entities. First, we create the WorldKG ontology, which is then used to describe the geographic entities in WorldKG. The steps involved in the WorldKG creation process are depicted in Figure~\ref{fig:pipeline}.

\subsection{WorldKG Ontology Creation}

The first part of the WorldKG creation process aims at creating the WorldKG ontology consisting of classes, properties, their relations, and links to the equivalent classes in Wikidata and DBpedia. This process consists of the following steps:

\begin{itemize}
    \item \textit{Scrape and filter key-value pairs:} First, we scrape the key-value pairs from the OSM map features which were introduced in Section~\ref{sec:ontology}. From these key-value pairs, we discard those that do not possess any class information. This concerns the key-value pairs categorized as \textit{additional attributes}, \textit{attributes} and \textit{additional properties} in OSM map features.
    \item \textit{Infer class hierarchy:} We use the keys to identify classes and key-value pairs to infer subclasses. If an individual value occurs with multiple keys, we manually specify a suitable subclass (e.g., for the key-value pairs \texttt{building=school} and \texttt{amenity=school}, we create classes \textit{BuildingSchool} and \textit{AmenitySchool}).
    \item \textit{Convert property and class names:} To adhere to established OWL naming conventions~\cite{owl2004}, we represent WorldKG classes in upper camel-case format and properties in lower camel-case format. 
    \item \textit{Schema alignment with Wikidata and DBpedia.} We establish \voc{owl}{equivalentClass} relationships to Wikidata and DBpedia ontologies through the schema alignment process described in Section~\ref{sec:ontology}.
\end{itemize}

\subsection{Knowledge Graph Creation}
After the creation of the WorldKG ontology, we now utilize this ontology to represent OSM nodes as geographic entities in WorldKG. This process includes the following steps:

\begin{itemize}
    \item \textit{Filter nodes with at least one tag:} As input, we retrieve all OSM nodes from the most recent OSM dumps\footnote{OSM dumps: \url{https://download.geofabrik.de/}} using the Osmium Python library\footnote{Osmium python library: \url{https://pypi.org/project/osmium/}}. 
    We filter out the nodes that do not contain any tags such as node:30519010\footnote{Filtered node: \url{https://www.openstreetmap.org/node/30519010}}. 
    These nodes are placeholders for ways and relations and are unlikely to be relevant for applications requiring node data.
\item \textit{Filter keys and values based on the WorldKG ontology:} Once we have collected the OSM nodes, we identify their classes and properties based on the WorldKG ontology and discard non-relevant tags and keys. From the OSM keys \textit{lat} and \textit{long}, we enrich the nodes with their geographic coordinates.
\item \textit{Create and validate triples:} Finally, we create RDF triples using the Python library \textit{RDFLib} and provide links to the corresponding resources in Wikidata and DBpedia. Geographic objects are represented as \voc{sf}{Point} objects pointing to the coordinates as \voc{geo}{WKTLiteral} literals. 
\end{itemize}

We provide an RDF dump of the geographic entities in WorldKG and its ontology. A SPARQL endpoint\footnote{WorldKG SPARQL endpoint: \url{http://www.worldkg.org/sparql}} using a Virtuoso triple store \cite{DBLP:series/sci/ErlingM09}
is set up to query WorldKG.

\section{WorldKG Characteristics \& Evaluation Results}
\label{sec:evaluation}
To illustrate the potential and quality of WorldKG, in this section, we present the statistics of the WorldKG and the evaluation results regarding the quality of the class alignment and type assertion.

\subsection{WorldKG Statistics}

As shown in Table \ref{tab:worldkgstat}, WorldKG contains more than 820 million triples associated with geographic data for 188 countries and seven continents. 
$33$ top-level classes were inferred from OSM keys, whereas the subclasses refer to the specific classes extracted from key-value pairs, as discussed in Section \ref{sec:ontology}.

\begin{table}[!h]
\caption{WorldKG knowledge graph statistics.}

\begin{tabular}{lr}
\toprule          

\textbf{Quantity} & \textbf{Count} \\ \midrule
Total triples             & 828,550,751      \\
Total entities             & 113,444,975      \\
Top-level classes             & 33    \\ 
Subclasses               & 1,143  \\ 
Unique properties         & 1,820  \\ 
Links to Wikidata classes & 40    \\ 
Links to DBpedia classes  & 21    \\ 
 \bottomrule
\end{tabular}
\label{tab:worldkgstat}
\end{table}

\subsection{Quality of the Class Alignment}

As reported in \cite{iswcNCA}, the NCA class alignment approach obtains matches with an average precision of 70\% and 90\% on the Wikidata and DBpedia knowledge graphs, respectively. 
As described in Section \ref{sec:schema-alignment}, we manually access the class alignments resulting from NCA and discard any wrong mappings to prevent the propagation of errors in WorldKG. 
By doing so, we obtain a class alignment precision of 100\%. 
This manual verification procedure does not affect the recall values. 
This way, the recall corresponds to the original NCA recall of 63\% and 81\% on Wikidata and DBpedia knowledge graphs, respectively, reported in \cite{iswcNCA}.

\subsection{Quality of the Type Assertion}

In this section, we assess the quality of type assertion in WorldKG regarding the Wikidata and DBpedia classes.
To this extent, we randomly select five classes from the DBpedia and the Wikidata ontologies mapped to WorldKG classes, respectively. For each of the resulting ten classes, we randomly select a sample of 100 WorldKG entities that are assigned to the respective class via \voc{rdf}{type} and \voc{owl}{equivalent\-Class}.  
Listing \ref{lst:evaluation_query} shows the SPARQL query used for the generation of a sample dataset for the Wikidata class Q556186 labeled ``mine''. 
For each of the resulting $1000$ entity-class pairs, we manually judge the correctness of the type assertion.
That way, we can estimate the accuracy of the type assertion in WorldKG.
The results are presented in Table \ref{tab:eval}.

 \begin{lstlisting}[float=ht,captionpos=b, caption= Query used to generate the sample set of 100 entities assigned to the Wikidata class Q556186 (``mine'')., label=lst:evaluation_query, frame=single]
SELECT ?id ?type ?osmid ?name 
WHERE {
  ?id rdf:type ?type  .
  ?id rdfs:label ?name.
  ?id wkgs:osmLink ?osmid.
  ?type owl:equivalentClass wd:Q556186.
} 
ORDER BY RAND() LIMIT 100


\end{lstlisting}

\begin{table*}[!h]
\vspace{5pt}
\caption{Evaluation results of the WorldKG type assertion regarding Wikidata and DBpedia classes.}
 \label{tab:eval}
\vspace{-5pt}
\begin{subtable}{\linewidth}
\centering
\caption{Wikidata}
\begin{tabular}{lrlrrrrr}
\toprule 
\textbf{WorldKG class} &  \textbf{WorldKG entities}   & \textbf{Wikidata class} & \textbf{Wikidata entities}  & \textbf{Correct}  & \textbf{Wrong} & \textbf{Non-verifiable} & \textbf{Accuracy} \\ \hline
Tomb    & 12849      & Q381885        &  3076    & 97 & 1 & 2                &   98.98\%       \\ 
Monument   & 44503  & Q4989906       &   23320    &   91   &    0     &    9            &   100.00\%       \\ 
Mineshaft & 8453   & Q556186        &    677     &  95   &       2  & 3 &   97.94\%       \\ 
BicycleRental & 40914 & Q61663696      &  1757   & 96 & 0 & 4                &    100.00\%     \\ 
TourismHotel & 204291 & Q27686         & 11152  & 97 & 0 & 3                   &  100.00\%    \\ 
\bottomrule
\end{tabular}
\label{tab:evalWiki}

\vspace{0.2 cm} 

\end{subtable}
\begin{subtable}{\linewidth}
\centering
\vspace{5pt}
\caption{DBpedia}
\begin{tabular}{lrlrrrrr}
\toprule
\textbf{WorldKG class}  & \textbf{WorldKG entities}   & \textbf{DBpedia class} & \textbf{DBpedia entities}  & \textbf{Correct}  & \textbf{Wrong} & \textbf{Non-verifiable} & \textbf{Accuracy} \\ \hline
\begin{tabular}[c]{@{}l@{}}ManMadeTower/\\ PowerTower\end{tabular} & 2769981 & Tower        & 2533      &  97 & 0  & 3 & 100.00\%      \\
City  & 10465   & City       &     22600     &     100 & 0 & 0       &      100.00\%    \\ 
Museum   & 46955  & Museum      & 7422 &  94 & 2 & 4                    &  97.92\%        \\ 
AmenitySchool & 424236 & School      &  31867 &      100 & 0 & 0             &  100.00\%        \\ 
CaveEntrance & 39525 & Cave         &  615 & 91 & 0 & 9                  &   100.00\%       \\ \bottomrule
\end{tabular}
\label{tab:evalDBp}

\end{subtable}

\end{table*}

Tables \ref{tab:evalWiki} and \ref{tab:evalDBp} present the evaluation results of the type assertion of WorldKG entities to the Wikidata and DBpedia ontology classes, respectively.
The \emph{correct} and \emph{wrong} columns indicate the number of correct or wrong KG classes assigned to individual entities, respectively.
The \emph{non-verifiable} column presents the number of cases in which we could not identify the correct class due to the lack of information about the entity on the web.
For instance, an OSM node tagged with \texttt{historic=monument}, with no further information available, can not be verified to actually be a monument\footnote{An example non-verfiable node: \url{https://www.openstreetmap.org/node/8752666922}}.
We exclude non-verifiable instances from our accuracy calculation.
As we can observe, the precise tag-to-class mappings in WorldKG facilitate a very high accuracy (between 97.9\% and 100\%) of type assertion regarding both Wikidata and DBpedia classes. The few cases of incorrectly assigned classes result from wrongly annotated instances in OSM. 

For all classes illustrated in Table~\ref{tab:evalWiki} and Table~\ref{tab:evalDBp} except of \voc{wkgs}{City}, the number of geographic entities in WorldKG is higher compared to Wikidata and DBpedia. Overall, as shown in Table~\ref{tab:worldkgComparison}, the number of geographic entities in WorldKG is two orders of magnitude higher than in Wikidata and DBpedia. 

Overall, the high accuracy class alignment of the WorldKG pipeline builds the foundation for the integration of OSM information into the linked open data cloud.
While OSM relies on the voluntarily contributed information, with no strict guarantees of correctness, WorldKG addresses this issue by only considering established tags defined in the OSM map feature list and therefore provides trustworthy high-quality information at scale. 

\begin{table}[!h]
\caption{Number of geographic entities in WorldKG, Wikidata and DBpedia}

\begin{tabular}{lr}
\toprule          

\textbf{Knowledge graph} & \textbf{Geographic entities} \\ \midrule
WorldKG              & 113,444,975      \\
Wikidata  & 8,621,058    \\ 
DBpedia   & 1,224,403   \\ 
 \bottomrule
\end{tabular}
\label{tab:worldkgComparison}
\end{table}

\section{Availability, Utility \& Sustainability}
\label{sec:availaility}

In this section, we describe how the WorldKG website and the data and code repositories ensure the availability, utility, and sustainability of WorldKG.

\subsection{Availability}
The WorldKG website\footnote{WorldKG Website: \url{http://www.worldkg.org/}} is publicly available. This website provides a description of WorldKG and a SPARQL endpoint for querying the WorldKG knowledge graph. In addition, the WorldKG website provides pointers to code and data:

\begin{itemize}
\item \textit{Code}: The code realizing the WorldKG creation process depicted in Figure~\ref{fig:pipeline} is available on GitHub\footnote{GitHub Link: \url{https://github.com/alishiba14/WorldKG-Knowledge-Graph}} under the MIT License\footnote{License for code: \url{https://opensource.org/licenses/MIT}}.
\item \textit{Data}: The WorldKG triples can be downloaded from a persistent URL\footnote{DOI for WorldKG data: \url{https://doi.org/10.5281/zenodo.4953986}} under the Open Data Commons Open Database License (ODbL)\footnote{License for dataset: \url{https://opendatacommons.org/licenses/odbl/}}. On the WorldKG website, we also made the manually created evaluation dataset of \voc{owl}{equi\-valent\-Class} mappings between the WorldKG ontology and the DBpedia/Wikidata ontologies available.
\end{itemize}

\subsection{Utility}

By following best practices in data publishing and open RDF W3C standard for modeling and interlinking the data, we envision World\-KG's establishment as part of the Linked Open Data Cloud. In detail, we ensure the utility of WorldKG via the following aspects:

\begin{itemize}

\item \textit{Documentation}: The WorldKG website provides a description of the data and the ontology. In addition, a selection of example SPARQL queries is given as an overview of potential usage scenarios and as a basis for creating new queries.

\item \textit{Data access}: WorldKG can be queried through its publicly available SPARQL endpoint that facilitates geographic queries with the option to download query results. Classes and resources of WorldKG can be looked up on the website, which also provides map visualizations pointing to the location of \textit{wkgs:WKGObject} entities.

\item \textit{Provenance}: Version 1.0 of WorldKG was extracted using OSM dumps from June 6, 2021\footnote{The OSM dumps were downloaded from \url{http://download.geofabrik.de/}.}. Among other metadata, this provenance information is provided as part of WorldKG using the VoID vocabulary~\cite{void2011}. Classes and instances in WorldKG are linked to Wikidata, DBpedia and OSM, where possible.
\end{itemize}

\subsection{Sustainability}

To keep WorldKG up-to-date with future releases of OSM that may further extend the coverage of real-world locations and account for potential transformations, we plan to publish new versions of WorldKG regularly.
We further plan to add additional features in the upcoming versions of WorldKG, 
including enriched entity descriptions and extended coverage of real-world entities through data fusion with other sources.

\section{Example Scenario}
\label{sec:example}
This section demonstrates the usage of WorldKG for Point-of-Interest (POI) recommendation through an example scenario.

\begin{lstlisting}[float=t, captionpos=b, caption=Example SPARQL query to retrieve the three closest restaurants to the Brandenburger Tor., label=lst:example_query, frame=single,showstringspaces=false]
PREFIX uom: 
<http://www.opengis.net/def/uom/OGC/1.0/>

SELECT ?closeObject ?restaurant 
 (bif:st_distance(?cWKT, ?fWKT, uom:metre)
  AS ?distance)
WHERE {
  ?poi rdfs:label "Brandenburger Tor".
  ?poi wkgs:spatialObject [
    geo:asWKT ?cWKT
  ] .
  ?closeObject rdf:type wkgs:Restaurant.
  ?closeObject rdfs:label ?restaurant.
  ?closeObject wkgs:spatialObject ?fGeom.
  ?fGeom geo:asWKT ?fWKT .
}
ORDER BY ASC(
  bif:st_distance(?cWKT, ?fWKT, uom:metre))
LIMIT 3

\end{lstlisting}

With the increased use of recreational and touristic applications, POI recommendation systems have gained increased attention \cite{DBLP:conf/www/LuoLL21,LIU2018183,Xie:2016:LGP:2983323.2983711}. Typically, the goal of a POI recommendation is to recommend a list of places to a user based on user-specific criteria, e.g., the user location and preferences. Knowledge graph, a machine-readable knowledge source supporting relational reasoning, can serve as a rich information source for POI recommendation \cite{DBLP:conf/sac/HalilajLRAD21}. However, cross-domain knowledge graphs often lack sufficient coverage of touristic POIs, such as restaurants. WorldKG fills this gap by providing means to retrieve POIs originated from OSM based on entity class labels.

Listing~\ref{lst:example_query} exemplifies a SPARQL query that returns the three closest restaurants (\voc{wkgs}{Restaurant}) for a given location (the Brandenburger Tor in Berlin, Germany). This query makes use of GeoSPARQL functions (i.e., \voc{bif}{st\_distance}) which are supported by the WorldKG SPARQL endpoint.

\begin{table}
\caption{Result of the example SPARQL query in Listing~\ref{lst:example_query}.}
\label{tab:example_result}
\begin{tabular}{lr}
\toprule
\textbf{Restaurant} & \textbf{Distance} \\ \hline
"Hopfingerbräu im Palais" & 0.128322 \\ 
"Restaurant Quarré" & 0.243953 \\ 
"Lorenz Adlon Esszimer" & 0.247478 \\ \bottomrule
\end{tabular}
\end{table}

\begin{figure}[!h]
    \centering
    
   \fbox{ \includegraphics[width = 0.45\textwidth]{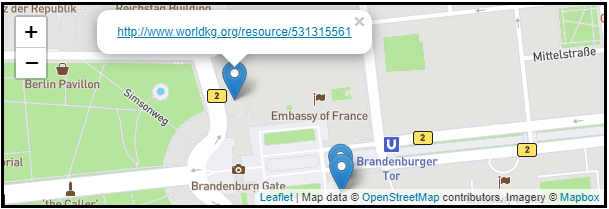}}
    \caption{Visualization of the three restaurants closest to the Brandenburger Tor returned by the query in Listing~\ref{lst:example_query}.}
    \label{fig:exampleMap}
\end{figure}

Table \ref{tab:example_result} shows the result of the example mentioned above after querying WorldKG, including the names of the restaurants and their distance to the Brandenburger Tor. Figure~\ref{fig:exampleMap} shows a screenshot taken from the result page of the WorldKG SPARQL endpoint, which is the visualization of returned restaurants on a map. This example demonstrates how POI applications can immediately benefit from the provision of geographic information in WorldKG.

\section{Related Work}
\label{sec:related}

In the following, we discuss existing KGs that include geographic entities, ontologies for geographic data and ontology alignment methods relevant for creating geographic KGs.

\subsection{KGs containing Geographic Entities}
There exist several dedicated geographic knowledge graphs as well as general-purpose knowledge graphs with geographic entities.
Wang et al. \cite{wang2019geographic} introduced GeoKG, a formalized geographic knowledge representation that complements the Attributive Language with Complements (ALC) description logic. Through case studies, the authors demonstrate that the GeoKG model can achieve more accurate and complete geographic knowledge representations compared to YAGO.

DBpedia \cite{articledbpedia} was one of the first well-established, cross-domain knowledge graphs. To represent geospatial data, DBpedia provides latitude and longitude values for various geographic entities. 
Similarly, Wikidata \cite{articlewikidata} represents coordinates of geospatial entities. However, both DBpedia and Wikidata only cover a small fraction of OSM locations \cite{TEMPELMEIER2021349}.
YAGO2geo \cite{karalis2019extending} is an extension of the YAGO knowledge graph that includes geospatial and temporal relations. YAGO2geo is created using OSM and reference geospatial datasets such as Greek Administrative Geography (GAG) and Global Administrative Areas dataset (GADM). YAGO2geo mainly focuses on administrative regions and reuses an existing ontology from the GAG dataset. 
EventKG \cite{gottschalk2019eventkg} is a knowledge graph that focuses on event-centric information and includes geographic entities relevant to historical events and their participants.  

In general, the knowledge graphs mentioned above lack coverage of geospatial entities and focus on certain entity types.
LinkedGeoData \cite{auer2009linkedgeodata}, on the other hand, converts OpenStreetMap data into an RDF knowledge graph. LinkedGeoData is based on a formal ontology created using tags and keys of OSM. it provides a simplified mapping between OSM data and classes and properties of other data sources. In contrast to WorldKG, LinkedGeoData uses a set of manually selected class mappings. Moreover, the latest available dumps of LinkedGeoData were released in 2015, and no links to Wikidata were provided.

\subsection{Ontologies for Geographic data}
There have been various approaches to build ontologies that cater to geographical data due to its unique structure. Sun et al. \cite{sun2019geospatial} have built a manual three-level ontology for geospatial data. Although the ontology they built can be reused, it is still incomplete and not assessed for quality.
With OSM being one of the most prominent sources of open geographic information, there have been approaches to build ontologies catering to the OSM data structure:
OSMOnto \cite{codescu2011osmonto} describes OSM tags in an ontology that provides few links to existing ontologies such as schema.org. Similar to WorldKG, OSMOnto is represented as a class hierarchy extracted from OSM keys and values.
Ballatore et al. \cite{ballatore2013geographic} developed the OSM semantic network by crawling OSM Wiki pages. The network can be used to compute the similarity between the concepts and also for geospatial retrieval of entities, among others. 
In contrast to these works, WorldKG ontology is created in an automated way and covers a variety of geographic classes. Thus, it is flexible towards OSM updates and not limited in its coverage of geographic entities.

\subsection{Ontology Alignment}
Ontology alignment (also known as ontology matching) aims to establish correspondences between the elements of different ontologies. 
The efforts to interlink open semantic datasets and to benchmark ontology alignment approaches have been driven by the W3C Semantic Web Education and Outreach (SWEO) Linking Open Data community project \cite{LOD}
and the Ontology Alignment Evaluation Initiative (OAEI) \cite{OAEI}
%
Ontology alignment is conducted at both the element-level and the structure-level \cite{Otero-CerdeiraRG15}.
The element-level alignment typically uses natural language descriptions of the ontology elements, such as labels and definitions. Element-level alignment adopts string similarity metrics such as edit distance.
Structure-level alignment exploits the similarity of the neighboring ontology elements, including the taxonomy structure, as well as shared instances \cite{NgoBT13}.
Element-level and structure-level alignments have also been adopted to align ontologies with relational data \cite{DemidovaON13} and tabular data \cite{ZhangB20}.
Jiménez-Ruiz et al. \cite{abs-2003-05370} divided the alignment task into independent, smaller sub-tasks, aiming to scale up to very large ontologies. 
Machine learning has been widely adopted for ontology alignment. In the GLUE architecture \cite{Doan2004}, semantic mappings are learned semi-automatically, while \cite{Nkisi-OrjiWMHH18} proposed a matching system that integrates string-based and semantic similarity features.
Recently, deep neural networks-based approaches have been used for ontology alignment and schema matching.
Proposed architectures include convolutional neural networks \cite{Bento2020OntologyMU}, representation learning \cite{Qiu2017KnowledgeEL}, and stacked autoencoders \cite{xiang-etal-2015-ersom}.
Until now, the lack of a well-defined ontology of OSM hindered the application of ontology alignment approaches to OSM data. 
WorldKG addresses this problem by providing alignments between OSM tags and knowledge graph classes. 
WorldKG builds upon the recently proposed Neural Class Alignment approach \cite{iswcNCA} that facilitates alignments between OSM tags and KG classes using a novel neural architecture.

\section{Conclusion}
\label{sec:conclusion}
In this paper, we presented WorldKG -- a new geographic knowledge graph that provides semantic representations of geographic entities in the OpenStreetMap dataset. 
The released WorldKG knowledge graph contains over 828 million triples of over 100 million entities spread across 1176 classes. 
Through manual quality assessment performed on randomly selected sample data, we observe that WorldKG contains high accuracy data. 
We make the data dump available and provide a SPARQL endpoint for accessing WorldKG. By following best practices for semantic data publishing, we ensure the availability and usability of the data and are committed to maintaining regular updates for sustainability. 
We believe that WorldKG has the potential to aid many applications and future research that consume geographic data.

\subsubsection*{\textbf{Acknowledgements}} This work was partially funded by DFG, German Research Foundation under ``WorldKG'' (424985896).

\balance
\bibliographystyle{ACM-Reference-Format}
\bibliography{references.bib}


\end{document}